\documentclass[twocolumn,aps,prl,showpacs]{revtex4}

\usepackage[dvips]{graphicx}
\usepackage{amsmath}

\begin{document}

\title{Observation of collective excitation \\of two individual atoms in the Rydberg blockade regime}

\author{A.~Ga\"{e}tan$^1$} 
\author{Y.~Miroshnychenko$^1$}
\author{T.~Wilk$^1$} 
\author{A.~Chotia$^2$} 
\author{M.~Viteau$^2$} 
\author{D.~Comparat$^2$} 
\author{P.~Pillet$^2$} 
\author{A.~Browaeys$^1$}
\author{P.~Grangier$^1$} 
\affiliation{$^1$Laboratoire Charles Fabry, Institut d'Optique, CNRS, Univ Paris-Sud, Campus Polytechnique, RD 128, 91127 Palaiseau cedex, France}
\affiliation{$^2$Laboratoire Aim\'{e} Cotton, CNRS, Univ Paris-Sud, B\^{a}timent 505, Campus d'Orsay, 91405 Orsay cedex, France}

\date{\today}

\begin{abstract}
The dipole blockade between Rydberg atoms has been proposed as a basic tool in quantum information processing with neutral atoms. Here we demonstrate experimentally the Rydberg blockade of two individual atoms separated by 4 $\mu$m. Moreover, we show that, in this regime, the single atom excitation is enhanced by a collective two-atom behavior associated with the excitation of an entangled state. This observation is a crucial  step  towards the deterministic manipulation of entanglement of two or more atoms using the Rydberg dipole interaction.
\end{abstract}


\maketitle

A large experimental effort is nowadays devoted  to the production of entanglement, that is quantum correlations,
between individual quantum objects such as atoms, ions, superconducting circuits, spins, or photons.
Entangled states are important in many areas of physics such as quantum information and quantum metrology,
the study of strongly correlated systems in many-body physics, and more fundamentally in the understanding of quantum physics.

There are several ways to engineer entanglement in a quantum system. Here, we focus on a method that relies on a
blockade mechanism where the strong interaction between different parts of a system prevents their simultaneous excitation by the same driving pulse. Single excitation is still possible, but it is delocalized over the whole system, and results in the production of an entangled state. This approach to entanglement is deterministic and can be used to realize quantum gates~\cite{Jaksch00} or to entangle mesoscopic ensembles, provided that the blockade is effective over the whole sample~\cite{Lukin01}. Blockade effects have been observed in systems where interactions are strong such as systems of electrons using the Coulomb force~\cite{Fulton87} or the Pauli effective interaction~\cite{Ono02}, as well as with photons and atoms coupled to an optical cavity~\cite{Birnbaum05}. Recently, atoms held in the ground state of the wells of an optical lattice have been shown to exhibit interaction blockade, due to s-wave collisions~\cite{Cheinet08}.

An alternative approach uses the comparatively strong interaction between two atoms excited to Rydberg states, which have very large dipole moments. This strong interaction gives rise to  the so-called  Rydberg blockade, which has been observed in  clouds of cold atoms~\cite{Tong04,Singer04,Afrousheh04,Cubel05,Vogt06,Ditzhuijzen08} as well as in a Bose condensate~\cite{Heidemann08}. A collective behavior associated with the blockade has been reported  in an ultra-cold atomic cloud~\cite{Heidemann07}. Recently, an experiment demonstrated the blockade between two atoms 10~$\mu$m apart,  by showing that when one atom is excited to a Rydberg state, the excitation of the second one is greatly suppressed~\cite{Urban08}.

In the present work, we study two individual atoms, held at a distance of $\sim 4$~$\mu$m by two optical tweezers. We demonstrate that under this condition, the atoms are in the Rydberg blockade regime since only  one atom can be excited. Furthermore, we show that the single atom excitation is enhanced by a collective two-atom behavior, associated with the production of a two-atom entangled state between the ground and Rydberg levels.

\begin{figure}
\begin{center}
\includegraphics[width=8cm]{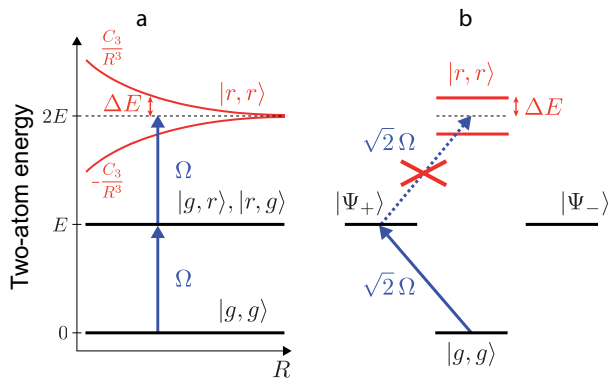}
\caption{(a) Principle of the Rydberg blockade between two atoms separated by a distance $R$. Two states $|g\rangle$ and $|r\rangle$ are coupled with Rabi frequency $\Omega$. When  the two atoms are in state $|r,r\rangle$ they interact strongly which leads to symmetrical energy shifts $\Delta E = \pm\, \frac{C_{3}}{R^3}$. When this shift becomes larger than  $\hbar \Omega$, the laser is out of resonance with the transition coupling the singly and doubly excited states, and only one atom at a time can be transferred to the Rydberg state. (b)  When the atoms are in the blockade regime, the state $|\Psi_{+}\rangle$, described in the text, is only coupled to the ground state $|g,g\rangle$ with a strength $\sqrt{2}\,  \Omega$ while the state $|\Psi_{-}\rangle$ is not coupled by the laser to the states $|g,g\rangle$ and $|r,r\rangle$. The atoms are therefore described by an effective two-level system.}\label{figure1}
\end{center}
\end{figure}

More specifically, let us consider the ground state  $|g\rangle$ and a Rydberg state $|r\rangle$ of an atom, separated by an energy $E$ (see figure~\ref{figure1}(a)) that can be coupled  by a laser. When two such atoms, $a$ and $b$, do not interact, the two-atom spectrum exhibits two transitions at the same frequency $E/\hbar$, connecting states $|g,g\rangle$ to $|r,g\rangle$ or $|g,r\rangle$, and then to  $|r,r\rangle$. This enables the simultaneous laser excitation of the two atoms to state $|r,r\rangle$. However, if the two atoms interact strongly when in state $|r,r\rangle$, this energy level is shifted  by an amount $\Delta E$ and the laser excitation can not bring the two atoms to  state $|r,r\rangle$. This ``blockade" effect has been suggested to
perform quantum gates and entangling operations~\cite{Jaksch00,Lukin01,Safranova03,Saffman05,Hyafil04}.

A fundamental consequence of this blockade  is that the atoms are excited in the entangled state $|\Psi_{+}\rangle = \frac{1}{\sqrt{2}}(e^{i {\bf k}\cdot {\bf r}_{a}}|r,g\rangle + e^{i {\bf k}\cdot {\bf r}_{b}}|g,r\rangle)$, where ${\bf r}_{a}$ and  ${\bf r}_{b}$ are  the positions of the two atoms, and ${\bf k}$ is related to the wavevectors of the exciting lasers (see below). More precisely, the laser excitation is described by the operator $\frac{\hbar \Omega}{2} (e^{i {\bf k}\cdot {\bf r}_{a}}|r,g\rangle\langle g,g| + e^{i {\bf k}\cdot {\bf r}_{b}}|g,r\rangle\langle g,g|+ {\rm complex\ conjugate})$~\cite{Dicke}. Here, $\Omega$ is the Rabi frequency characterizing the coupling between the laser and \textit{one} atom. It is then convenient to use as a basis the two entangled states $|\Psi_{\pm}\rangle = \frac{1}{\sqrt{2}}(e^{i {\bf k}\cdot {\bf r}_{a}}|r,g\rangle \pm e^{i {\bf k}\cdot {\bf r}_{b}}|g,r\rangle)$, so that the state $|\Psi_{-}\rangle$ is not coupled to the ground state, while the state $|\Psi_{+}\rangle$ is coupled with an effective Rabi frequency $\sqrt{2}\, \Omega$. In the blockade regime, where the state $|r,r\rangle$ is out of resonance, the two atoms are therefore described by an effective two-level system involving collective states $|g,g\rangle$ and $|\Psi_{+}\rangle$ coupled with a strength of $\sqrt{2}\, \Omega$, as shown in figure~\ref{figure1}(b). Therefore it is predicted that the atoms are excited into an entangled state containing only one excited atom, with a probability which oscillates $\sqrt{2}$ times faster than the probability to excite one atom when it is alone.

In our experiment, we excite two rubidium 87 atoms to the Rydberg state $|r\rangle = |58d_{3/2}, F =3, M_{F}=3\rangle$, with the internuclear axis aligned along the quantization axis. The state $58d_{3/2}$ was chosen because of the existence of a quasi-degeneracy between the two-atom states $(58d_{3/2}, 58d_{3/2})$ and $(60p_{1/2}, 56f_{5/2})$, also called a F\"orster resonance~\cite{Walker08}, that enhances the interaction strength. Using the procedure outlined in~\cite{Walker08}, one finds that the dipole-dipole interaction lifts the degeneracy and leads to two potential curves $\pm\, \frac{C_{3}}{R^3}$, as represented in figure~\ref{figure1}(a). For our particular geometry, we calculated $C_{3} \approx h \times 3200\ \rm{MHz}\cdot \mu\rm{m}^3$. Therefore, for a distance  $R=4$ $\mu$m, the interaction energy between the two atoms is $\Delta E \approx h\times 50$ MHz.

\begin{figure}
\includegraphics[width=8cm]{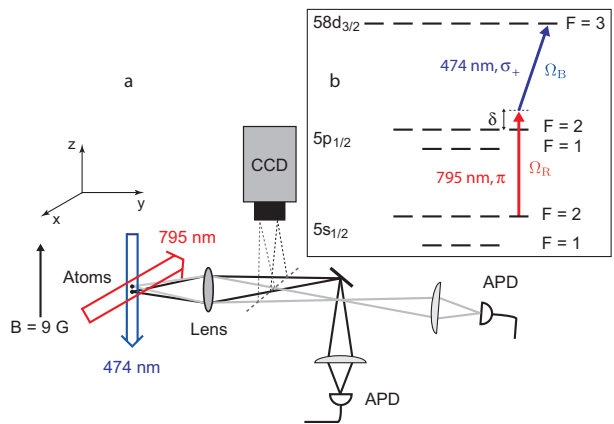}
\caption{(a) Experimental setup. Two rubidium 87 atoms are trapped in two tightly focused dipole traps (not shown). The quantization axis is set by a 9 G magnetic field. The fluorescence of each trap is collected on two separated  avalanche photodiodes (APD) in single photon counting mode. We can also image the two atoms on a CCD camera, which allows to measure the distance between the atoms.
(b) The atoms are excited to the Rydberg state by a two-photon transition. An infrared
laser at 795 nm propagating along the $x$-axis, with  $\pi$-polarization, couples  states $|5s_{1/2}, F=2, M_{F}=2\rangle$ and $|5p_{1/2}, F=2, M_{F}=2\rangle$. The laser is detuned to the blue of this transition by $\delta = 400$ MHz. The second laser at 474 nm is $\sigma_{+}$-polarized, coupling the states  $|5p_{1/2}, F=2, M_{F}=2\rangle$ and $|58d_{3/2}, F =3, M_{F}=3\rangle$, and propagates along the $z$-axis.
The infrared laser has a waist of $\approx 130 \ \mu$m and a power of 7 mW. The 30 mW blue laser is focused on 25~$\mu$m. From a light-shift measurement, we get a Rabi frequency $\Omega_{\rm R} \approx 2\pi \times 260$ MHz.
The two-photon Rabi frequency, measured experimentally (see figure~\ref{figure3}), is given by $\Omega = \frac{\Omega_{\rm R}\Omega_{\rm B}}{2 \delta}$, leading to $\Omega_{\rm B} \approx 2\pi \times 21$ MHz.} \label{figure2}
\end{figure}

The two single rubidium 87 atoms are confined in two independent optical dipole traps, as shown in figure~\ref{figure2}(a). The traps are formed in the focal plane of the same large numerical aperture lens~\cite{Schlosser01}. Each trap has a waist of 0.9 $\mu$m and a depth of 0.5 mK. The distance between the two traps can be varied between 3 and 20 $\mu$m with a precision of 0.5 $\mu$m. The axis between the two traps is aligned with the magnetic field materializing the quantization axis. The traps are loaded from the cold atomic cloud of an optical molasses. We collect the fluorescence emitted by the atoms, induced by the cooling lasers,
on two separated single photon counters. A high fluorescence level is the signature of the presence of an atom in the trap. The experimental sequence, described hereafter, is triggered on the presence of the atoms.

We prepare the two atoms in the hyperfine state $|g\rangle = |5s_{1/2}, F=2, M_{F}=2\rangle$ by a 600~$\mu$s
optical pumping phase~\cite{opticalpumping}. The efficiency of the pumping is $\sim 90$\%, which we attribute to an imperfect polarization of the pumping beams. We then excite the atoms to the Rydberg state $|r\rangle = |58d_{3/2}, F =3, M_{F}=3\rangle$ by a two-photon transition, as represented in figure~\ref{figure2}(b). The intermediate state   $|5p_{1/2}, F = 2, M_{F}=2\rangle$  is connected to the ground state by a laser detuned by 400 MHz to the blue of the
795 nm transition to avoid populating the intermediate state. The second laser connects the $5p_{1/2}$ to the $58d_{3/2}$ states and has a wavelength of 474 nm. Both laser beams illuminate the two atoms. During the excitation ($<500$ ns), the dipole trap is turned off to avoid an extra light-shift on the atoms. We finally detect the excitation to the Rydberg state through the loss of the atom when the dipole trap is turned back on. When an atom is in a Rydberg state, it is not trapped by the optical potential any longer. Due to its residual velocity, on the order of 10 cm/s,  the atom leaves the trapping region in less than 10 $\mu$s, well before the atom has the time to decay to the ground state~\cite{detection}.

\begin{figure}
\includegraphics[width=8cm]{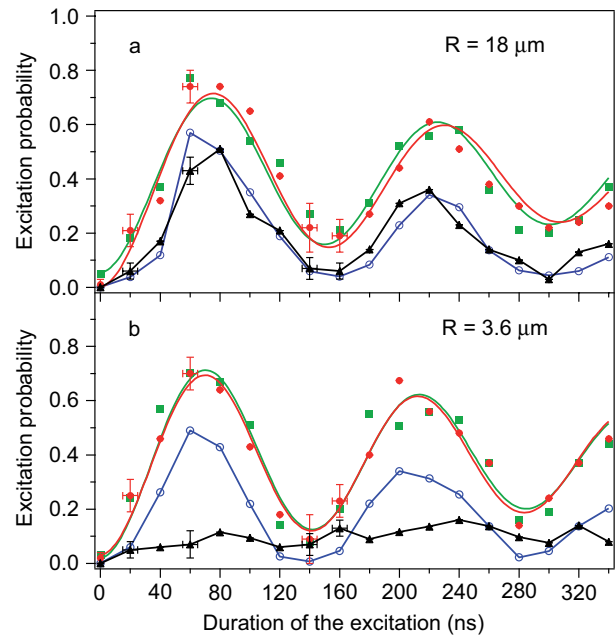}
\caption{Rydberg excitation of two atoms. In both figures, the red circles and the green squares represent the
probability to excite atom $a$ and atom $b$ respectively, when the other atom is absent. We fit the data by
the function $A -B e^{-\frac{t}{\tau}} \cos \Omega t$, shown
as plain line. The error bars on the data are the RMS statistical error on the measured
probability, as well as the error in the estimation of the pulse duration.
The blue empty circles are the product of the probabilities
to excite atom $a$ and atom $b$ when the other one is absent.
The triangles are the probability
to excite the two atoms simultaneously when they are driven
by the same pulse.
(a) Atoms separated by 18   $\mu$m. The frequencies of the Rabi oscillations are 6.5 MHz and 6.4 MHz
for atom $a$ and $b$ respectively. The agreement between the triangles and the blue circles indicates that
the atoms do not interact.
(b) Blockade of the Rydberg excitation when the two atoms are separated by 3.6  $\mu$m.
Due to the  interaction between the atoms,
this double excitation is greatly suppressed. } \label{figure3}
\end{figure}

In a first  experiment, we placed the two traps at a distance of $18 \pm 0.5 $ $\mu$m and repeated the excitation sequence 100 times starting each time with newly trapped atoms. For each sequence, we measured for each of the atoms whether it was lost or recaptured at the end of the sequence. We then calculated the probability to excite the atom in the Rydberg state, which is equal to the probability to lose the atom. When only one of the two traps is filled with an atom, the excitation probability exhibits Rabi oscillations between the ground state and the Rydberg state, as shown in figure~\ref{figure3}(a). A fit to the data yields a two-photon Rabi frequency $\Omega\approx 2\pi \times 7$ MHz which is in  agreement with the measured waists and powers of the lasers. The  decay of the amplitude of the fringes is explained by the frequency fluctuations ($\sim 1$ MHz) as well as  shot-to-shot  intensity fluctuations of the lasers ($\sim 5$\%), which results into a jitter in the two-photon resonance frequency. We attribute the maximum excitation
probability of $\sim 80$\% to this decay and to the imperfect optical pumping of the atoms in the Zeeman state $|5s_{1/2}, F=2,M_{F}=2\rangle$~\cite{frequency_comment}.

We then repeated the sequence with two atoms trapped at the same time and we measured the probability to excite the two atoms with the same laser pulse. The results are represented in figure~\ref{figure3}(a) by the triangles. We compared this probability with the probability to excite simultaneously two non-interacting atoms, which should be equal to the product of the probabilities to excite each atom independently, measured previously. The blue circles in figure~\ref{figure3}(a) represent this product, calculated from the data for each independent atom. The agreement between the two curves shows that the two atoms, when separated by 18 $\mu$m, behave independently and therefore have a negligible interaction. This result agrees with the theory since the blockade becomes effective at a distance between the atoms for which the interaction shift $\Delta E$ is equal to the linewidth of the excitation pulse, on the order of the Rabi frequency $\Omega$. For our particular choice of the Rydberg state, this yields  $R \approx 8\ \mu$m.

In a second experiment, we set the distance between the two traps to $3.6 \pm 0.5 $ $\mu$m, in a regime where we expected the blockade to occur, and we repeated the previous experiment. Once again we measured the probability to excite one atom when the other one is absent and got the one-atom Rabi oscillations. When two atoms were trapped, we measured the probability to excite the  two atoms simultaneously, as shown in figure~\ref{figure3}(b) by the triangles. We observed that the simultaneous excitation of the two atoms is greatly suppressed with respect to the case where the atoms are far apart. This suppression is the signature of the blockade regime. We also observed that the probability of simultaneous excitation of the two atoms is not completely cancelled. This fact may be explained by the existence of extra potential curves coming from imperfect control of the state of the atoms and leading to smaller interaction energies~\cite{Walker08}. This imperfect control can be due to stray electric fields,  imperfect polarizations of the lasers, and random positions of the atoms in their trap  (see last paragraph) meaning that the inter-nuclear axis is not always perfectly aligned with the quantization axis.

\begin{figure}
\includegraphics[width=8cm]{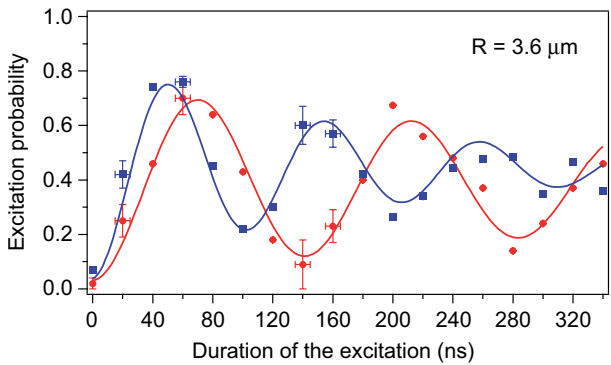}
\caption{Collective excitation of two atoms separated by 3.6~$\mu$m. The circles represent
the probability to excite atom $a$ when atom $b$ is absent (same curve as in figure~\ref{figure3}(b)).
A fit to the data yields a frequency of this Rabi oscillations $\Omega/ 2\pi =
7.0\pm 0.2$ MHz. The error comes from the fit and corresponds to one standard deviation.
The squares represent the probability to excite only one atom when the two atoms are
trapped and are exposed to the same excitation pulse. The fit gives an oscillation frequency
$\Omega' /2\pi= 9.7\pm 0.2$ MHz. The ratio of the oscillation frequencies is $1.38 \pm 0.03$, close to the value
$\sqrt{2}$ expected for the collective oscillation of two atoms between $|g,g\rangle$ and $|\Psi_{+}\rangle$. }\label{figure4}
\end{figure}

We now come to the direct observation of collective one-atom excitation in the blockade regime, i.e.  with two atoms separated by  $R=3.6\ \mu$m. Figure~\ref{figure4} shows  the probability to excite only one of the two atoms as a function of the duration of the excitation pulse, together with the probability to excite only one atom when the other dipole trap is empty. The two probabilities oscillate with different frequencies, which ratio is $1.38\pm 0.03$ (the error corresponds to one standard deviation). This value is compatible with the ratio $\sqrt{2}$ that we expect when the two atoms interact strongly enough to be in the blockade regime. As explained previously, the oscillation of the probability to excite only one atom at a frequency $\sqrt{2}\, \Omega$ is the signature that the two-atom system oscillates between the state $|g,g\rangle$ and the entangled state $|\Psi_{+}\rangle = \frac{1}{\sqrt{2}}(e^{i {\bf k}\cdot {\bf r}_{a}}|r,g\rangle + e^{i {\bf k}\cdot {\bf r}_{b}}|g,r\rangle)$, where ${\bf k} = {\bf k_{R}+k_{B}} $ is the sum of the wavevectors
of the two lasers involved in the two-photon transition.

Finally, we analyze the influence of the atoms' motion on this entangled state. We measured a temperature of the atoms in their trap of 70 $\mu$K~\cite{Tuchendler08}. This leads to amplitudes of the motion of $\pm 800$ nm in the longitudinal ($y$) direction and $\pm 200 $ nm in the radial ($x$ and $z$) direction of the traps~\cite{oscillationfrequency}. Since the fastest oscillation period is 13 $\mu$s and the excitation time is
on the order of hundred nanoseconds, the motion of the atoms is frozen during the excitation. The temperature only results in a dispersion of the positions of the atoms from shot to shot. Therefore the relative
phase $\phi = {\bf k}\cdot ({\bf r}_{a}-{\bf r}_{b})$ between the two components of the superposition is constant during the excitation, but varies randomly from shot to shot over more than $2\pi$. This creates an effective decoherence mechanism for the state $|\Psi_{+}\rangle$, which would prevent the direct observation of the entanglement. However, this fluctuating phase can be erased in the following way: one  first couples one hyperfine ground  state $|0\rangle$ to a Rydberg state $|r\rangle$, producing the state  $|\Psi_{+}\rangle=\frac{1}{\sqrt{2}}(|r,0\rangle + e^{i\phi}|0,r\rangle)$ as described in this paper. Then a second pulse, applied before the atoms move, couples $|r\rangle$ to a second hyperfine ground state $|1\rangle$.
If the wavevectors of the two excitations are the same, the phase during the second step cancels the phase of the first excitation. The resulting entangled state is therefore  $\frac{1}{\sqrt{2}}(|1,0\rangle + |0,1\rangle)$, which involves long-lived atomic qubits.

In conclusion, we have demonstrated that the strong interaction between two atoms separated by a few micrometers prevents the simultaneous excitation to Rydberg states. In addition, we have observed the collective oscillation of the pair of atoms between the ground and the singly-excited states, with a Rabi frequency increased by $\sqrt{2}$, which is consistent with the deterministic excitation of the entangled state
$\frac{1}{\sqrt{2}}(|r,g\rangle + e^{i\phi}|g,r\rangle)$. This result indicates that we control the physical mechanism needed to deterministically entangle two atoms in two hyperfine ground states. Combined with our abilities to manipulate the state of a single atom~\cite{Jones07}, to keep and to transport its quantum state~\cite{Beugnon07}, our system is well adapted to applications of the Rydberg blockade in quantum information processing.

\begin{acknowledgments}
We thank Mark Saffman, Thad Walker, Robin C\^ot\'e and Thomas Gallagher for enlightening discussions. We thank Ivan Liu for theoretical support  and Thomas Puppe for technical assistance with the lasers. We thank Charles Evellin for calculations and experimental assistance, as well as Lucie Servant. We acknowledge support from the EU through the IP SCALA, IARPA and IFRAF. A.~G. is supported by a DGA fellowship. Y.~M. and T.~W.  are supported by the IFRAF.
\end{acknowledgments}


\begin{thebibliography}{30}
\bibitem{oscillationfrequency} The oscillation frequencies of the atoms in the trap are 77~kHz in the radial ($x$ and $z$) direction and 16~kHz in the longitudinal ($y$) direction.
\bibitem{opticalpumping} We overlapped temporally a pumping $\sigma_{+}$-polarized laser tuned to the transition between  $(5s_{1/2}, F=2)$ and  $(5p_{3/2}, F=2)$ with a repumper beam resonant with the $(5s_{1/2}, F=1)$ to  $(5p_{3/2}, F=2)$ transition.
\bibitem{detection} The radiative decay of the Rydberg state $58d_{3/2}$ is 500~$\mu$s, while the lifetime induced by the blackbody radiation is 50~$\mu$s, calculated from T. F. Gallagher, \textit{Rydberg atoms}, Cambridge (1994).
\bibitem{frequency_comment} For our parameters, the frequency of the two-photon transition depends mainly on the light-shift of the $F=2$ ground state induced by the 795 nm laser. This light-shift is different for different Zeeman sublevels due to different Clebsch-Gordan coefficients. Therefore, the two-photon excitation resonant with the transition between  $|5s_{1/2}, F=2, M_{F}=2\rangle$ and $|58d_{3/2}, F =3, M_{F}=3\rangle$ does not excite the other Zeeman sublevels of the  $(5s_{1/2},F=2)$ manifold.
\end{thebibliography}
\end{document}